\documentclass[sigconf,nonacm]{acmart}
\acmConference{ }{ }{ }

\AtBeginDocument{%
  }
\usepackage[most]{tcolorbox}
\usepackage{hyperref}
\usepackage{array}
\usepackage{float}
\usepackage{tabularx}
\renewcommand\footnotetextcopyrightpermission[1]{}
\setcopyright{none}
\newtcolorbox{designbox}{
  enhanced,
  colback=yellow!8,          
  boxrule=0pt,               
  frame hidden,
  borderline west={3pt}{0pt}{red!70!black}, 
  arc=1mm,                   
  left=4mm, right=4mm, top=2mm, bottom=2mm
}

\begin{document}
\title{A Comparative Analysis of Peer Support in Forum-based and Chat-based Mental Health Communities: Technical-Structural-Functional Model of Social Support}

\author{Han Li}
\affiliation{%
  \institution{Cornell University}
  \city{Ithaca}
  \state{NY}
  \country{USA}
}
\email{hl2564@cornell.edu}

\begin{abstract}
Online support communities have become vital spaces offering varied forms of support to individuals facing mental health challenges. Despite the proliferation of platforms with distinct technical structures, little is known about how these features shape support dynamics and the socio-technical mechanisms at play. This study introduces a technical–structural–functional model of social support and systematically compares communication network structures and support types in 20 forum-based and 20 chat-based mental health communities. Using supervised machine learning and social network analysis, we find that forum-based communities foster more informational and emotional support, whereas chat-based communities promote greater companionship. These patterns were partially explained by network structure: higher in-degree centralization in forums accounted for the prevalence of informational support, while decentralized reply patterns in chat groups accounted for more companionship. These findings extend the structural–functional model of support to online contexts and provide actionable guidance for designing support communities that align technical structures with users’ needs. 
\end{abstract}



\maketitle

\section{Introduction}
Mental disorders constitute a major contributor to the global burden of disease, with significant implications for public health and economic development. In China, they are among the two leading causes of healthy life loss \cite{Zhou_2019}. According to a 2019 national survey, an estimated 11\% to 15\% of the population exhibits symptoms indicative of mild to moderate mental health issues, while 2\% to 3\% may be affected by severe mental illness. Despite the growing demand for mental health care, access to timely and adequate treatment remains limited in China, particularly in rural areas \cite{xiang_2018}. This disparity has given rise to the emergence of online peer support communities across different social media platforms, which serve as accessible and scalable sources of information, emotional support, and shared experience for individuals coping with mental health challenges. Rooted in peer-based interaction, such communities have demonstrated clear value, with evidence showing that online peer support can enhance psychosocial well-being \cite{allison_2021,rains_2025}, foster mental health learning and individual empowerment \cite{jones_2011, zhang_2018}, and reduce the likelihood of suicidal thoughts \cite{munmun_2017}. 

Since the early days of online communities such as Usenet, the form and structure of online peer support communities have continued to evolve alongside advances in digital communication technologies. Today, they are broadly organized by two predominant formats: forum-based communities, which support structured, asynchronous exchanges centered around persistent discussion threads; and chat-based communities, which enable synchronous interactions in a more fluid and loosely organized conversational flow \cite{li_2021}. Despite a wealth of HCI research has respectively examined support exchanges and community dynamics within forum-based and chat-based peer support communities \cite{Gui_2017,Rzai_2020,Yadav_2022}, a critical gap remains in systematically comparing and analyzing whether, and in what ways, their distinct interface characteristics and technical structures shape communication patterns and the nature of peer support. Framing online communities as socio-technical systems, Preece and De Souza \cite{souza_2004,preece_2001} proposed a foundational framework that highlights sociability and usability as two critical qualitative factors influencing the success of online communities. These two dimensions — referring to the social interactions among members and the design of the system interface — jointly shape users' experiences and engagement. More importantly, the system interface serves not merely as a medium, but plays a vital role in shaping the way members interact with each other and group outcomes \cite{li_2021}.

At the core of online peer support communities is the dynamic exchange of social support. Prior research has shown that individuals turn to these spaces to both seek and provide a rich spectrum of support, ranging from informational resources, to emotional validation and reassurance, and at times, the simple comfort of companionship and shared presence. The nature of the support exchanged not only reflects the multifaceted needs of people suffering from mental health challenges, but also influences the psychosocial benefits members gain from the community \cite{vlahovic_2014}, as well as the patterns of participation and the enduring vitality of the community itself \cite{wang_2017,wang_2015}. Although researchers have repeatedly observed that format or mode of Computer-Mediated Communication (CMC) shapes the types of social support exchanged within online communities \cite{Chuang_2012, xie_2008, chuang_2014, green_hamann_2011}, this line of inquiry remains limited in important ways. Existing studies have either focused narrowly on variations within forum-based platforms (e.g., forum, journal, and notes) \cite{Chuang_2012,chuang_2014}, or have relied primarily on qualitative data to explore distinctions between forum-based and chat-based systems \cite{xie_2008,green_hamann_2011}. Despite these valuable insights, a systematic comparison of the types of support exchanged across these two dominant socio-technical systems of online peer support communities is still lacking. Moreover, the mechanisms through which technical structures shape support dynamics remain under-theorized and empirically underexplored. 

In this paper, we address two central questions: 1) Whether, and in what ways, the types of social support exchanged differ between forum-based and chat-based communities; and 2) how distinct communication patterns help explain such differences. To this end, we analyze over 182,778 online messages drawn from 40 peer support communities focused on mental health topics such as depression, anxiety, and bipolar disorder. These communities were sampled from four widely used Chinese social media platforms — Baidu Tieba,  Douban Group, WeChat and QQ, with 10 communities from each platform. As such, our comparison is structured between communities (N=20) from two chat-based platforms (i.e., WeChat and QQ) and communities (N=20) hosted on two forum-based platforms (i.e., Baidu Tieba and Douban Group).   

The contributions of this study are threefold. First, our findings illuminate systematic differences in both the structural and functional dimensions of support, advancing theoretical understanding of the mechanisms that shape peer support dynamics within chat-based and forum-based communities. Second, we introduce the \textit{technical-structural-functional model of social support}, extending existing theoretical frameworks by integrating the role of platform design and technical characteristics in mediating support exchanges. This extended model offers a more holistic account of how peer support unfolds in digitally mediated settings. Third, we discuss the design implications of our work for digital platforms aimed at supporting individuals with diverse support needs, highlighting how technical features can be navigated and tailored to gratify users and address the trade-offs between informational and companionship values of peer support communities. 

\section{Related Work}

\subsection{Mental health Support in Online Communities}
Online peer support communities have been a long-standing focus of interest in HCI research, particularly as spaces where individuals voluntarily seek and provide support around mental health challenges. These communities represent a distinctive form of digitally mediated social context, where users initiate and engage in social interactions that are often emotionally charged, context-sensitive, and high-stakes. Such exchanges carry profound implications not only for individual well-being \cite{allison_2021}, but also for the cultivation of supportive community norms and collective resilience \cite{li_2021,oleary_2017}. Accordingly, the design and governance of these communities have drawn increasing scrutiny within HCI, prompting calls for more evidence-based, human-centered approaches to foster high-quality social support, effective intervention and responsive moderation\cite{oleary_2017}.

A dominant thread in HCI's investigation of mental health support in the context of online communities focuses on characterizing the \textit{content} of social interactions. This includes work that maps the types of support exchanged across user populations and contexts  (e.g.,  \cite{Gui_2017,siddiqui_2023,progga_2023}) and studies that unpack the linguistic features and interaction patterns associated with beneficial outcomes (e.g., \cite{sharma_2018,wang_2012}). Parallel to this, design efforts have explored mechanisms and best practices to facilitate support seeking and provision, such as matching algorithms that connect users with potential helpers \cite{Fang_2022}, as well as writing tools that enhance message quality \cite{peng_2020,shin_2022}.

Compared to a major focus on \textit{content}, the \textit{context} that mediates these interactions received comparatively less attention. Notably, a small body of research has examined cultural contexts as influential external forces shaping users' support needs and interpersonal dynamics within peer support communities \cite{pendse_2019,li_2024, zhang_2018}. Beyond culture, technical context constitutes another critical yet understudied dimension that may shape the nature, structure, and quality of social support. For example, Chuang and Yang \cite{Chuang_2012} conducted a content analysis of an alcoholism support group on MedHelp to examine how three distinct CMC formats—discussion board, blog, and notes, shaped the forms of support exchanged. They found that emotional support appeared more frequently in blog and notes than in discussion board. A follow-up study \cite{chuang_2014} further revealed that blog and notes primarily served relational functions, facilitating the maintenance of interpersonal ties, whereas discussion board was predominantly used for informational exchanges. These findings echo earlier work by \cite{xie_2008} and \cite{green_hamann_2011}, which similarly highlight the role of CMC format and modality in structuring support behaviors. 

While these studies offer valuable empirical evidence that technical contexts shape support dynamics, much of this work has remained descriptive and focused on single platform or community. What remains missing is a systematic theoretical model that links platform affordances — such as synchronicity and conversational structures — with specific mechanisms of supportive behavior. Without such a model, the relationship between technical context and the dynamics of social support remains conceptually fragmented and empirically underdeveloped within HCI research. To address this gap, the present study analyzes the nature of social support exchanged in forum-based and chat-based online peer support communities, aiming to disentangle the specific ways in which differences in synchronicity and message organization shape the types of support exchanges. In the sections that follow, we review foundational theories of social support, present a systematic comparison of forum-based and chat-based online communities, and examine relevant communication network mechanisms to inform the development of a technical-structural-functional model of social support in digitally mediated contexts. 

\subsection{Social Support: Concept, Type, and Approach}
Social support, originally conceptualized to address the critical question of how personal social networks protect individuals from illness and mortality, has long been recognized as a vital coping resource and a key buffering mechanism. It helps individuals navigate uncertainty, manage psychological distress and cognitive dissonance, and recover from stressful life events \cite{albrecht_2003,cohen_1985}. Approaching social support as a communicative process, \cite{burleson_1994} identified three core dimensions that define supportive interactions: the content of messages exchanged during support episodes; the conversational contexts in which these messages are produced, interpreted, and evaluated; and the broader impact of support communication on psychosocial dynamics. These dimensions continue to offer a foundational lens for understanding how social support operates and the outcomes it generates in both offline and online settings. 

The functional approach to social support centers on enacted support, defined as "specific lines of communication behavior enacted with the intent of benefiting or helping one another" \cite{burleson_2002}. A central focus of this line of research on enacted support lies in the categorization of support types \cite{rains_2015}, grounded in the premise that different forms of support are optimally effective under particular stress conditions \cite{cutrona_1992, Fang_2022}. A widely cited framework proposed by \cite{cutrona_1992} identifies distinct categories: informational support, emotional support, network support, esteem support, and tangible support. While this taxonomy has been extensively applied in offline contexts, studies of online environments, especially peer support communities have revealed a different pattern. As Rains and Wright observe \cite{rains_2016}, informational and emotional support are particularly predominant in digital settings, whereas network, esteem, and tangible support appear less frequently. In addition, scholars have increasingly drawn attention to companionship as a distinct and often overlooked form of support in online spaces \cite{huang_2014,liu_2020}. \cite{rook_1987} defines companionship as shared leisure activities pursued primarily for intrinsic enjoyment, distinguishing it from more problem-focused forms such as informational or emotional support. In the latter, whether one is offering advice, providing information,or conveying empathy and comfort, the core function is to help support seekers cope with specific personal difficulties, stressors, or distress. Companionship, by contrast, is sought and provided "for its own sake", serving a more phatic and hedonic goal of mutual leisure and recreation \cite{rook_1987}. 

Rather than aiming to solve or soothe, companionship emphasizes the relational and experiential rewards of presence and connection. As such, acknowledging companionship is essential to developing a more comprehensive understanding of supportive communication \cite{barrera_1983}. In response to these evolving insights, a growing number of studies have adopted a three-category framework for analyzing social support in online exchanges: informational support, emotional support, and companionship support \cite{huang_2014, xie_2008, wang_2017}. Our work builds on this tradition, employing the three-category classification to examine the types of support exchanged in forum-based and chat-based mental health communities. 

Another important strand of research in supportive communication has advanced understanding of the social network resources that underlie the provision and exchange of social support. Thoits \cite{thoits_1982} introduced the concept of “social support system” to describe subgroups within an individual’s broader social network that offer socio-emotional and tangible assistance. Subsequent studies have examined the extent to which individuals are embedded in networks of supportive ties, often operationalizing social embeddedness through measures such as contact frequency and relationship strength \cite{wills_2012}. However, this approach has been critiqued for its narrow focus on dyadic interactions, which tends to obscure the broader structural properties of networks that shape how support flows through a community \cite{meng_2016, zhu_2013}. As such, scholars have increasingly sought to map the general communication networks through which support is exchanged. For example, HCI research on newsgroups revealed that online depression support communities tended to exhibit denser and more dynamic communication networks compared to other types of forums, suggesting that the structure of communication network can vary depending on a community's purpose \cite{muncer_2000}. Recent work has further examined how specific network structures are associated with different types of support. In a social network analysis of interactions within a mental health forum, it was found that network density and cohesion varied based on whether users exchanged informational or emotional support \cite{chang_2009}. Similarly, \cite{pfeil_2009} observed that empathic support networks tended to form tighter connections than those centered on factual or informational exchanges, highlighting how different forms of support are embedded within distinct structures of communication networks.  

The structural approach to social support offers a vital complement to the functional view by foregrounding the role of social ties and network characteristics that enable access to supportive resources. Given that network structures can shape, constrain, or amplify the forms of support that enacted within them, an integrative structural-to-functional support model has been proposed to capture structural patterns and the communicative functions they serve \cite{lin_1999}. The model theorizes how structural characteristics such as size, density, and connectivity condition individuals' access to functional support resources, which in turn impact their psychological and social well-being. Though originally developed to explain support processes and outcomes in offline networks, the model has proven valuable for analyzing the dynamics of support in digitally networked settings. For example, Meng et al. \cite{meng_2016} investigated how network structures, specifically network brokerage and closure within online health communities shaped the reception of different types of social support. Their findings revealed that brokerage was positively associated with the receipt of informational and network support, while closure was linked to the receipt of emotional and esteem support. Expanding on this work, a subsequent study examined social support exchanges in both online and offline contexts among cancer patients, revealing that structural characteristics of each type of network had a distinct influence on patient emotional well-being. Specifically, larger offline networks were negatively associated with emotional well-being, due to their tendency to increase informational support. In contrast, larger online networks were positively associated with emotional well-being by facilitating greater access to emotional support \cite{meng_2021}. 

Our work is situated centrally in this evolving body of research, where we introduce a \textit{technical-structural-functional model of social support} to examine how communication technologies mediate the processes of supportive exchanges. A meta-analysis on social support in online health contexts underscores the critical importance of studying technological variation across communication media and its effects on the quality and type of social support \cite{rains_2015}. Because communication technologies actively configure the affordances for interaction, they may play a constitutive role in shaping both how support is distributed and what forms it takes within online communities.   

\subsection{Forum-based versus Chat-based Online Communities}
\label{sec:socio-technical system}
From early Usenet groups and IRC channels to contemporary platforms like Reddit, Discord, and WhatsApp, the format of online communities have continually evolved in tandem with technological advancements. Herring (2002) distinguished between "older modes" and "newer modes" of CMC, offering a useful framework for understanding how technical modes shape interaction dynamics. According to this typology, older modes of CMC — such as Usenet newsgroups, forums, BBS,and mailing lists—are typically asynchronous and persistent, meaning that interactions unfold with temporal delays and are persistently archived for later retrieval. In contrast, newer modes of CMC such as voice chat, instant messaging tend to be synchronous and ephemeral. Recent research has drawn a parallel distinction in the context of online communities. Specifically, common formats of online communities can be broadly categorized into forum-based communities, which exhibit the characteristics of older CMC modes, and chat-based communities, which align more closely with the properties of newer modes \cite{li_2021}. 

A foundational technical distinction between forum-based and chat-based online communities lies in their degree of synchronicity, which reflects how communication is temporally coordinated and structured \cite{albrecht_1987}. Forum-based communities operate on a highly asynchronous model, where users are not expected to be co-present in real time. This temporal flexibility allows participants to compose, edit, and respond to messages at their own pace, often over the course of hours, days, or even weeks \cite{levin_2006}. \cite{kalman_2006} compared reply latency across three asynchronous platforms — email, university forums, and Google Answers and found that the average response time was 28.76 hours on email, 23.52 hours on university forums, and 1.58 hours on Google Answers. 

In contrast, chat-based communities are marked by high synchronicity. Originating from early technologies like IRC and evolving into contemporary platforms such as Discord and WhatsApp, these systems facilitate rapid, real-time exchanges. In synchronous settings, messages function not only as conveyors of content but as signals of social presence. Responsiveness is performative: quick replies and message frequency communicate engagement, while delays or silence can be easily interpreted as absence or ignorance \cite{donath_1999}. As \cite{bays_1998} notes, prolonged pauses in synchronous contexts may be perceived as socially inappropriate, making temporal immediacy an implicit norm. Seufert et al. \citeyearpar{seufert_2015}, analyzed over 220,000 messages from 271 WhatsApp group chats and found that nearly 60\% of messages received a reply within one minute, and over 80\% within fifteen minutes—highlighting the immediacy characteristic of chat-based environments. 

Synchronicity also intersects with a second key temporal dimension: the persistence of conversations. Forum-based systems tend to exhibit high persistence, with topic-based threads archived for long durations, enabling asynchronous users to revisit, review, and contribute to prior discussions at any time \cite{nonnecke_2001}. This persistence supports the accumulation of shared knowledge and facilitates longitudinal engagement within the community. Conversely, chat-based communities typically exhibit low persistence, prioritizing immediacy over durability. Many real-time messaging systems do not retain messages beyond a short time frame. For example, early IRC servers lacked the ability to store conversations for more than an hour \cite{bays_1998}. In modern chat interfaces, messages appear in a linear, time-stamped stream and are gradually pushed out of view as new messages arrive. While temporary scrolling enables limited retrospective access, deleted messages or chatroom removals can result in permanent loss of content. This difference in temporal structure is considered a critical group-level feature that shapes the effectiveness and nature of social support exchanges. Building on a meta-analysis of 28 computer-medaited support communities, \cite{rains_2009} concluded that patients participating in interventions that incorporated synchronous communication technologies reported greater increases in perceived social support compared to those limited to asynchronous platforms. 

Another salient distinction between forum-based and chat-based communities lies in the organization of messages. Forum-based communities typically structure posts and comments into discrete “threads”, with replies hierarchically nested under the initiating post. This threaded representation affords a higher degree of conversational structure and coherence, enabling participants to follow discussions more easily and positioning the thread initiator— often the support-seeker—at the center of the interaction. In contrast, chat-based communities present messages in a chronological stream through a linear interface. This time-based arrangement results in a less structured conversational flow. As noted by \cite{kim_2021}, the structures of online discussions can influence a range of social processes in online groups, such as the diversity of viewpoints expressed and the deliberative quality of exchanges. 

\subsection{Communication Network}
As posited by the structural-to-functional support model \cite{meng_2021,lin_1999}, the structural patterns of communication networks operate as key mechanisms accounting for support dynamics. Studies of various online communities have devoted considerable attention to map and characterize the communication networks that arise within them. Drawing on social network analysis, Pfeil and Zaphiris \cite{pfeil_2009} examined how the structural features of peer-support networks related to the content of member exchanges. Their findings revealed a clear association: denser, more tightly knit networks fostered richer empathetic exchanges, whereas factual support tended to emerge within sparser networks characterized by greater distance between members. Extending this line of work, Owen et al. \cite{owen_2016} conducted a 12-week social network intervention to evaluate the network structures developed across four distinct platforms: forum, chatroom, email, and blog. Among them, chatroom exhibited the greatest network density, while forum was the most diffuse. In addition, chatroom surpassed forum in clustering and average degree, reflecting a more interconnected network environment. This body of empirical evidence suggests that the formation and evolution of communication networks within online communities rely critically on ongoing interactions among members, which can be shaped by the technical structures that underlie these communities. 

\begin{figure}[!ht]
  \centering
  \includegraphics[width=\linewidth]{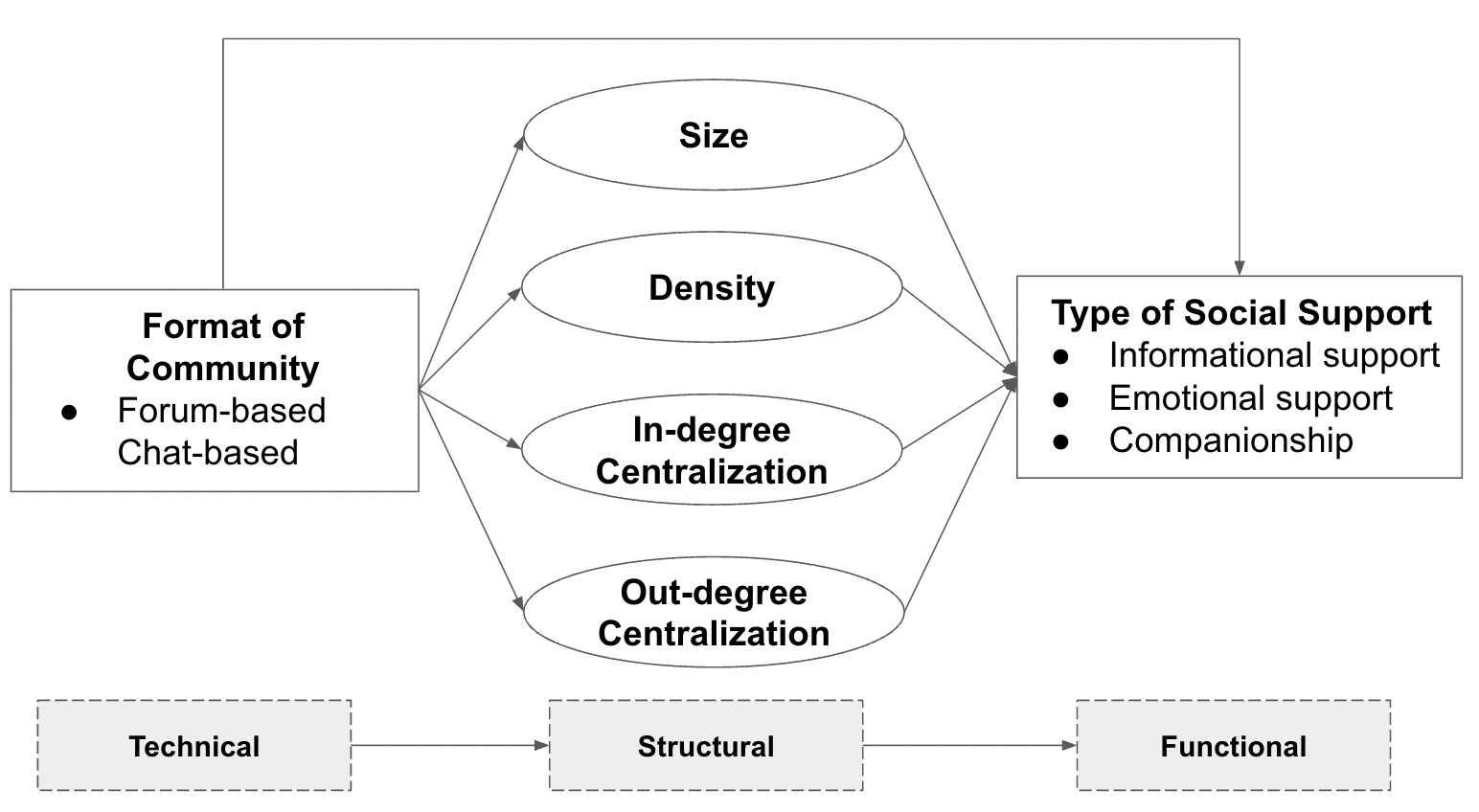}
  \caption{Analytical Framework: Technical-Structural-Functional Model of Social Support}
  \Description{}
  \label{fig0}
\end{figure}

In this study, we focus on three network-level structural indicators-\textit{network size}, \textit{network density}, and \textit{network centralization (in-degree and out-degree)}- that may vary between forum-based and chat-based communities due to the synchronicity and persistence of communication, as well as the degree to which conversations are structured and centralized. Figure.\ref{fig0} illustrates the technical-structural-functional model of social support introduced and validated in this study.   

\section{Method}

\subsection{Research Sites}
To enable meaningful comparisons between form-based and chat-based online communities, we collected one month of publicly available data from 40 Chinese online mental health communities across four major social media platforms: Baidu Tieba, Douban Group, WeChat and QQ. Baidu Tieba and Douban Group primarily host forum-based communities, whereas WeChat and QQ support chat-based communities. To identify relevant communities with a focus on mental health, we conducted keyword searches using six terms associated with significant mental disorders: \textit{depression}, \textit{anxiety disorder}, \textit{bipolar disorder}, \textit{eating disorder}, and \textit{personality disorder}. This search yielded over 300 candidate communities across the four platforms. After excluding communities with very low activity levels (i.e. fewer than 10 available threads in forums or fewer than 100 messages in chat groups during the one month observation period), we selected 10 communities from each platform, yielding a final sample of 40 online mental health communities for analysis. As shown in Table \ref{tab:table1}, the thematic composition of communities on forum-based and chat-based platforms is broadly comparable. However, chat-based platforms include slightly more communities focused on mood disorders, whereas forum-based platforms contain more communities centered on personality disorders, despite our efforts to keep the thematic focus as similar as possible across platforms. 

\begin{table}
\centering
\begin{tabular}{
>{\centering\arraybackslash}p{0.4\linewidth}
>{\centering\arraybackslash}p{0.15\linewidth}
>{\centering\arraybackslash}p{0.15\linewidth}
}
\toprule
Community thematic focus & Forum-based platforms & Chat-based platforms \\
\midrule
Mood disorders (depressive disorder + bipolar) & 10 & 13 \\
Anxiety-related disorders (anxiety disorder, social phobia, OCD) & 6 & 5 \\
Personality disorders (BPD, AvPD) & 3 & 1 \\
Eating disorder & 1 & 1 \\
\midrule
Total & 20 & 20 \\
\bottomrule
\end{tabular}
\caption{Thematic focus of the 40 online communities across platforms}
\end{table}

The study received review and approval from the Institutional Review Board at a research-intensive university prior to data collection. For communities on the two forum-based platforms, data were collected from publicly accessible pages. For communities on chat-based platforms, we identified potential groups through QR codes that had been publicly shared on open-access forum platforms (e.g., Tieba and Douban), where users promote chat groups related to mental health topics. Using these QR codes, the first author joined chat groups that were explicitly described as peer support spaces. Although membership needed to be approved by a community administrator, access was generally open: applications were routinely accepted without requiring justification or personal information. No deception was involved in the joining process. We deliberately chose not to join chat groups when further authentication steps were required. While we acknowledge that chat platforms are often used for private communication, the public availability of group invitations and the minimal entry requirements suggest that these communities are somewhat more open than typical private chats and may be better understood as a relatively semi-public space. 

While obtaining explicit individual informed consent was often not feasible to conduct observational studies in real-world social media contexts, we implemented the following measures to uphold ethical standard, following best practices for ethical Internet research \cite{franzke_2020}. Specifically, 1) all data were anonymized at the time of collection, with usernames replaced by pseudonyms; 2) We did not reproduce any verbatim user quotes in the paper; 3) all data were stored securely on encrypted drives with access restricted to the research team; and 4) no private or public messages were sent to group members at any stage of the study.

\subsection{Data Collection}
We employed Python-based web scrapers to collect all posts and comments from Baidu Tieba and Douban Group. For WeChat and QQ, we used off-the-shelf message management tools to download the complete set of accessible group chat messages. In addition to the textual content, we collected available metadata, including timestamps, post or message IDs, and usernames. In total, our dataset comprised 3,338 threads containing 25,481 posts and comments from the 20 forum-based communities, and 157,197 messages from 20 chat-based communities. Across all platforms, 8,269 users contributed to the 182,778 textual entries included in our analysis. A independent-samples t-test showed that there was no significant difference in participation size (i.e. the number of members who actively contributed to threads or conversations during the one-month observation period) between chat-based and forum-based platforms (t=-.29, df=36.01, \textit{p}=.77).   

Because chat-based communities present messages in an unstructured, time-ordered stream rather than in structured, topic-based threads as seen in forum-based communities, we implemented a manual annotation process to reconstruct coherent conversational threads. To this end, we recruited and trained four research assistants to manually extract topic-based conversations and annotate reply relationships (i.e. the intended recipient of each non-initiating message). The annotation process involved two primary tasks. The first was thread identification, in which coders grouped topically related messages posted within a 12-hour window into distinct conversational threads. The second was target identification, where coders labeled the intended recipient of each message based on explicit user mentions (e.g., use of the @ symbol) or inferred from the surrounding conversational context. This procedure resulted in the reconstruction of 7,275 conversational threads within chat-based communities. 

The primary objective of this study was to examine the relationships among format of community, communication network, and the type of social support exchanged within forum-based and chat-based online communities. To operationalize these constructs, our analysis proceeded in three stages. First, we trained and validated machine learning models to automatically classify the types of social support expressed in each text entry; Second, we constructed communication networks based on identified reply relationships among active participants, and computed key structural features — including network size, density, and concentration — that characterize the topology of each communication network. Third, we employed multilevel modeling techniques to assess whether the types of social support exchanged differed by format of the community (i.e., forum-based versus chat-based), and how structural characteristics of the communication network account for the observed differences in support dynamics. A detailed description of each stage of the analysis is presented in Section \ref{measures}. 

\subsection{Measures}
\label{measures}
\subsubsection{Automated Annotation of Support Type}
\label{automated annotation}
Given the large volume of data, we employed supervised machine learning techniques to automatically classify the types of social support expressed in the textual content, which has been well validated in previous research on online peer support communities \cite{wang_2015, wang_2017, huang_2010}. Drawing on a theory-informed framework, we adopted a three-category typology of social support, classifying each message into one of the following types: informational support, emotional support, or companionship support. Our classification pipeline consisted of five key steps: (1) construction of human-labeled ground-truth; (2) feature curation; (3) algorithm selection; (4) model training and evaluation, and (5) final prediction. Each of these components is described in detail in the sections that follow.    

(1) Construction of human-labeled ground-truth. To construct a ground-truth dataset for feature curation and model evaluation, we proportionally sampled 5,000 textual entries from the 40 online communities. Two trained coders annotated each entry using a well-defined coding schema for identifying types of social support. Each textual entry was single-coded, such that coders assigned one primary support type to each message. To assess inter-coder reliability, both coders independently annotated the same set of 1,000 entries, achieving a Cohen’s k over 0.9, indicating high agreement (see Appendix for our coding schema, example, and a confusion matrix for inter-rater reliability). The remaining 4,000 entries were evenly divided and independently coded. From the fully annotated corpus, we selected 1,000 entries from each support category (i.e., informational support, emotional support, and companionship support) to construct a balanced dataset of 3,000 labeled entries, which served as the ground-truth for training and evaluating the machine learning models.  

(2) Feature Curation. Drawing on prior research in online support classification \cite{wang_2015,wang_2017,huang_2010}, we extracted four categories of linguistic and thematic features to capture the multifaceted characteristics of the textual content for use in the machine learning models. Specifically, the feature set included: LIWC (Linguistic Inquiry and Word Count) features, LDA topic features, mental health lexicon features, and sentiment features.    

a. LIWC features. LIWC is a widely adopted text analysis program designed to quantify the cognitive, affective, and structural properties of language. It has been validated as a reliable predictor in studies of social support detection \cite{wang_2015}. In this study, we employed LIWC 2015 in conjunction with the SC-LIWC (Simplified Chinese LIWC dictionary \cite{huang_liwc_2012}) to extract psycholinguistic features from each textual entry in our ground-truth dataset. This process yielded 87 features per text, with each feature corresponding to a predefined linguistic or psychological category (e.g., affect, cognition, social processes). Together, these features provided a comprehensive linguistic profile of the user-generated content, enabling the model to capture subtle cues related to the type of support expressed. 

b. LDA topic features. While LIWC features capture a broad range of psycholinguistic attributes in language use, they are less effective at representing thematic content — a critical dimension for distinguishing types of social support. For example, prior research has found that themes of encouragement and comfort were frequently associated with emotional support, whereas discussions related to medication and treatment often co-occurred with informational support \cite{buis_2008}. To capture such thematic distinctions, we applied LDA topic modeling to uncover the underlying themes present in the corpus. Following practices in existing studies \cite{wang_2015,wang_2017}, we randomly sampled 60,000 textual entries from the full dataset to train the LDA model, specifying 20 latent topics. Standard pre-processing steps were performed, including the removal of numbers, punctuation, and stop words. To improve computational efficiency and reduce noise, we retained only words that appeared at least five times in the corpus. We ran 500 iterations of Gibbs sampling to ensure model stability and convergence. For each of the 20 identified topics, we extracted the top 500 words with the highest probability of association, resulting in topic-word distributions that were used as input features for the machine learning classifiers. Additionally, we manually reviewed the top words for each topic and assigned interpretive labels to describe their thematic focus.  

c. Mental health lexicon features. To enhance the domain specificity of our classification models, we developed a custom Chinese mental health lexicon by aggregating frequently occurring terms from across the 40 online communities in our dataset. The lexicon was organized into five categories: disorder names (e.g. “depression”, “bipolar”), common medications (e.g., “amitriptyline”, “clomipramine”), symptoms and side effects (e.g., “insomnia”, “anxiety”), treatment methods (e.g., “mindfulness”, “hypnosis”), and medical resources (including names of hospitals, clinics, and mental health professionals). Following the same procedure used for LDA topic features, we calculated term frequencies for each annotated message by computing the overlap between words in the message and entries in the lexicon. These frequency values were then incorporated as input features in the machine learning models, capturing the presence and salience of mental health-related language. 

d. Sentiment features. Prior research has shown that while mental health lexicon features are particularly effective in identifying informational support, sentiment-related features are better suited for detecting emotional support, as they capture affective cues embedded in language \cite{wang_2017}. As such, we included sentiment features into the model. Specifically, we utilized the HowNet Sentiment Dictionary, a widely used Chinese lexical resource comprising 836 positive sentiment words and 1,254 negative sentiment words \cite{hownet_2002}. For each message, we calculated the frequency of positive and negative sentiment words, generating two sentiment features that reflect the emotional valence of the text. 

Based on the four feature categories described above, we constructed a comprehensive textual feature set comprising 114 features in total. This feature set was designed to capture both the linguistic form and thematic content of each message, as well as domain-specific and affective cues relevant to social support classification. 

(3) Algorithm selection. We used the \textit{caret} package in RStudio to implement a series of machine learning models, drawing on four commonly adopted algorithms for classification tasks: k-nearest neighbors (KNN), decision trees, support vector machines (SVM), and random forests. To avoid overfitting and mitigate potential redundancy within the feature set, we also evaluated the performance of various feature subset combinations. This allowed us to assess not only the performance of each algorithm but also the contribution of different feature types to classification performance. The final model was selected based on the optimal pairing of algorithm and feature set, balancing predictive accuracy and model robustness.  

(4) Model training and evaluation. To train and evaluate the classification models, we randomly split the ground-truth corpus of 3,000 annotated messages into a training set (2,400 messages; 80\%) and a test set (600 messages; 20\%). Model training was conducted using 10-fold cross-validation on the training set. In this procedure, the training data were partitioned into ten equal subsets; in each iteration, one subset was used as a validation set while the remaining nine were used for model training. This approach ensured robust performance estimation and minimized overfitting. The final performance of each classifier was assessed using the held-out test set, with overall classification accuracy serving as the primary evaluation metric. The model that achieved the highest accuracy on the test set was selected as the final classifier, which was then applied to the remaining unlabeled messages in the dataset for automatic support type classification.

\subsubsection{Structural Characteristics of Communication Network}
To map the structural characteristics of communication networks within online communities, we constructed reply-based directed networks representing interactions among active participants on a weekly basis. The unit of analysis was the community-week. For each community and each week, we created a separate directed network in which nodes represent all users who posted at least once during that week and edges represent at least one reply from one user to another within the same week. Because networks were constructed independently for each week, the same user could appear as node in multiple community-week networks if they were active across several weeks. However our analyses do not treat users or threads as independent observations. Instead, the unit of analysis in all models is aggregated at the community-week level. All edges are directed, pointing from the replier to the original poster. 

In total, we constructed 160 community-week reply networks (40 communities X 4 weeks). To analyze the structure of these networks, we examined three key metrics: network size, network density, and network centralization (in-degree and out-degree). All network construction and analysis were conducted using the \textit{igraph} and \textit{sna} packages in RStudio. 

\textit{Network size.} As a fundamental metric in social network analysis, network size quantifies the number of unique users actively engaged in supportive exchanges within a given community-week. This metric reflects the scope of participation in each network. Given that the distribution of network size exhibited a long-tailed pattern, we applied a log\textsubscript{10} transformation to normalize the raw values for analysis.

\textit{Network Density.} Network density captures the overall level of connectedness among participants by measuring the proportion of observed edges (i.e., reply relationships) relative to the total number of possible edges in the network. A density value of 1 indicates a fully connected network, where every user replied to every other user at least once, whereas a density of 0 denotes a complete absence of interaction. This metric offers insight into how sparsely communication is distributed within each community-week.

\textit{Network Centralization.} We used network centralization to assess inequality in interaction patterns, that is, how unevenly communication is distributed among participants. In directed networks, centralization can be assessed along two dimensions: out-degree and in-degree. Out-degree centralization indicates the extent to which a small number of users initiate the majority of replies, while in-degree centralization reflects whether a few individuals receive a disproportionate share of replies. High centralization values suggest that interaction is concentrated among a few dominant users, whereas low values indicate a more egalitarian, decentralized communication structure. 

Specifically, we began by calculating each participant’s out-degree (i.e. the number of messages sent) and in-degree (i.e. the number of messages received) within each thread. To quantify the centralization of these distributions, we employed the \textit{ineq} package in RStudio to compute the Gini coefficient for both out-degree and in-degree values at the thread level, which has been widely adopted as a robust indicator of inequality \cite{lambert_1993}. The overall centralization of a given communication network was derived by averaging the Gini coefficients across all threads within the network. Both out-degree and in-degree centralization scores range from 0 and 1, where values closer to 0 indicate a more even distribution of message exchange, and values approaching 1 signify a highly skewed distribution of interaction dominated by a few users. 

\subsection{Data Analysis}
Given the hierarchical structure of the dataset, where weekly interaction data are nested within communities, we employed multilevel structural equation modeling to account for the nested nature of the data and appropriately model within- and between-community variance. Specifically, we tested three key relationships:

(1) the association between format of community (forum-based versus chat-based) and the proportion of different types of social support exchanged;

(2) the association between format of community and four structural characteristics of the communication networks: network size, network density, out-degree centralization, and in-degree centralization;

(3) the role of these network structural characteristics in explaining associations between format of community and the types of social support exchanged.

All models were constructed and estimated using Mplus 8, which supports multilevel mediation models for complex data structures.

\section{Findings}
\subsection{Differences in Support Type between Forum-based and Chat-based Communities}
We began by examining the distribution of support types across forum-based and chat-based communities, using supervised machine learning to automatically classify the support expressed in each message. As a baseline, we implemented a Naive Bayes classifier with Bag-of-Words features, which achieved an accuracy of 0.46 on the test set. To improve performance, we developed 16 models combining various feature sets and algorithms, as detailed in \ref{automated annotation}. 

The best-performing model was a Random Forest trained on a combination of LIWC, LDA topic, and mental health lexicon features, which achieved a training accuracy of 0.761. Subsequent feature importance analysis identified 75 highly predictive features; retraining the model using only this refined feature set yielded a modest improvement in accuracy to 0.765. When applied to the held-out test set, the optimized model achieved an overall accuracy of 0.83, substantially outperforming the baseline Naive Bayes classifier (0.46). Performance by support type was highest for companionship (0.86), followed by informational (0.82) and emotional support (0.8). These results align with prior studies on English-language datasets, which typically report classification accuracies between 0.7 and 0.9 \cite{wang_2015,wang_2017,huang_2010}. The optimized model was subsequently deployed to classify all remaining unlabeled messages in the dataset.

\begin{figure}[!ht]
  \centering
  \includegraphics[width=\linewidth]{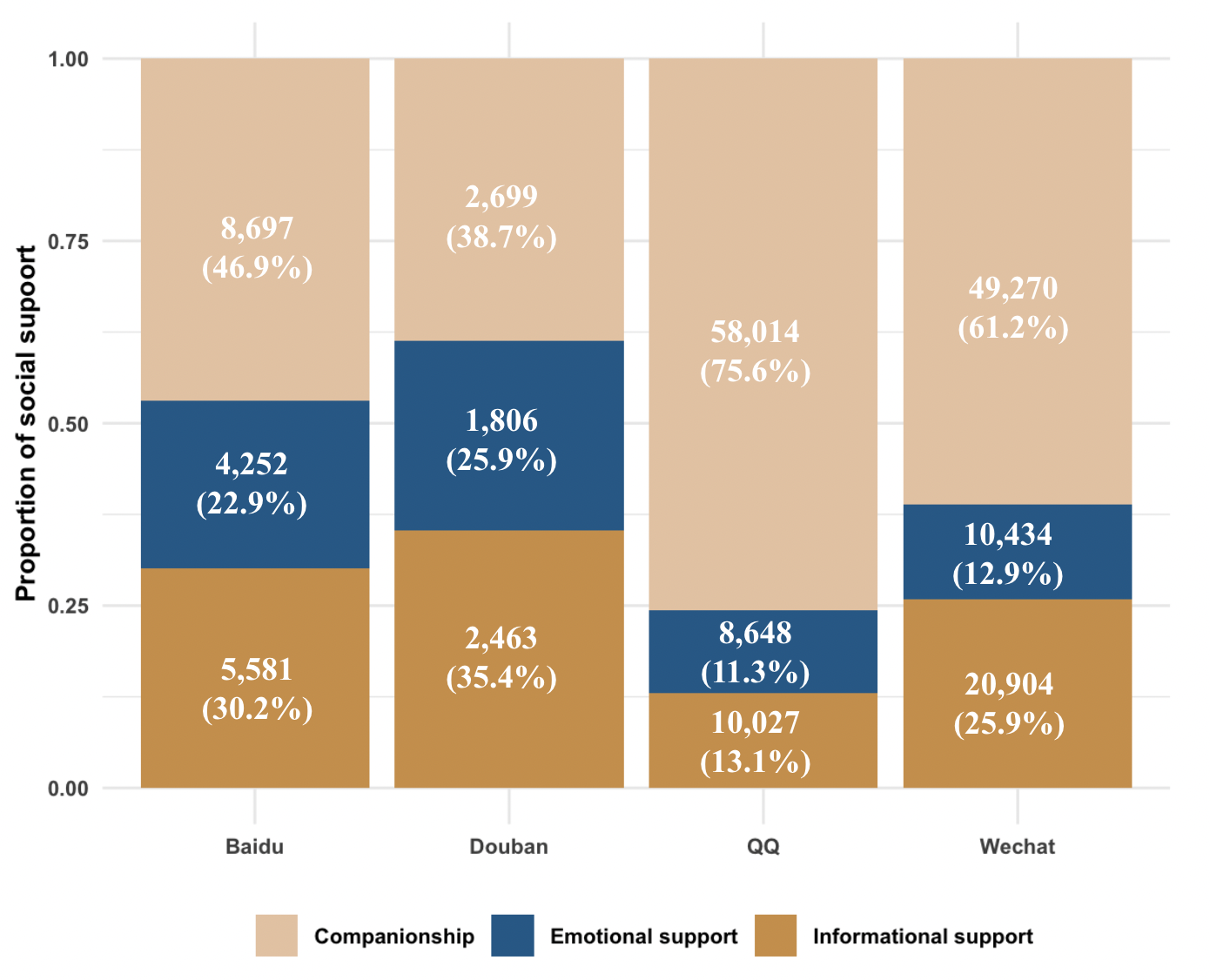}
  \caption{Proportions of informational, emotional and companionship support across four social media platforms}
  \label{fig1}
  \Description{}
\end{figure}

Figure \ref{fig1} presents the proportions of the three support types — emotional, informational, and companionship — across the four platforms. Overall, companionship emerged as the most prevalent, followed by informational and emotional support. A Shapiro-Wilk test indicated that only the proportion of companionship met normality assumptions, while those for informational and emotional support did not. Accordingly, we conducted an independent samples t-test for companionship and Wilcoxon rank-sum tests for the other two categories.

\begin{figure}[!ht]
  \centering
  \includegraphics[width=\linewidth]{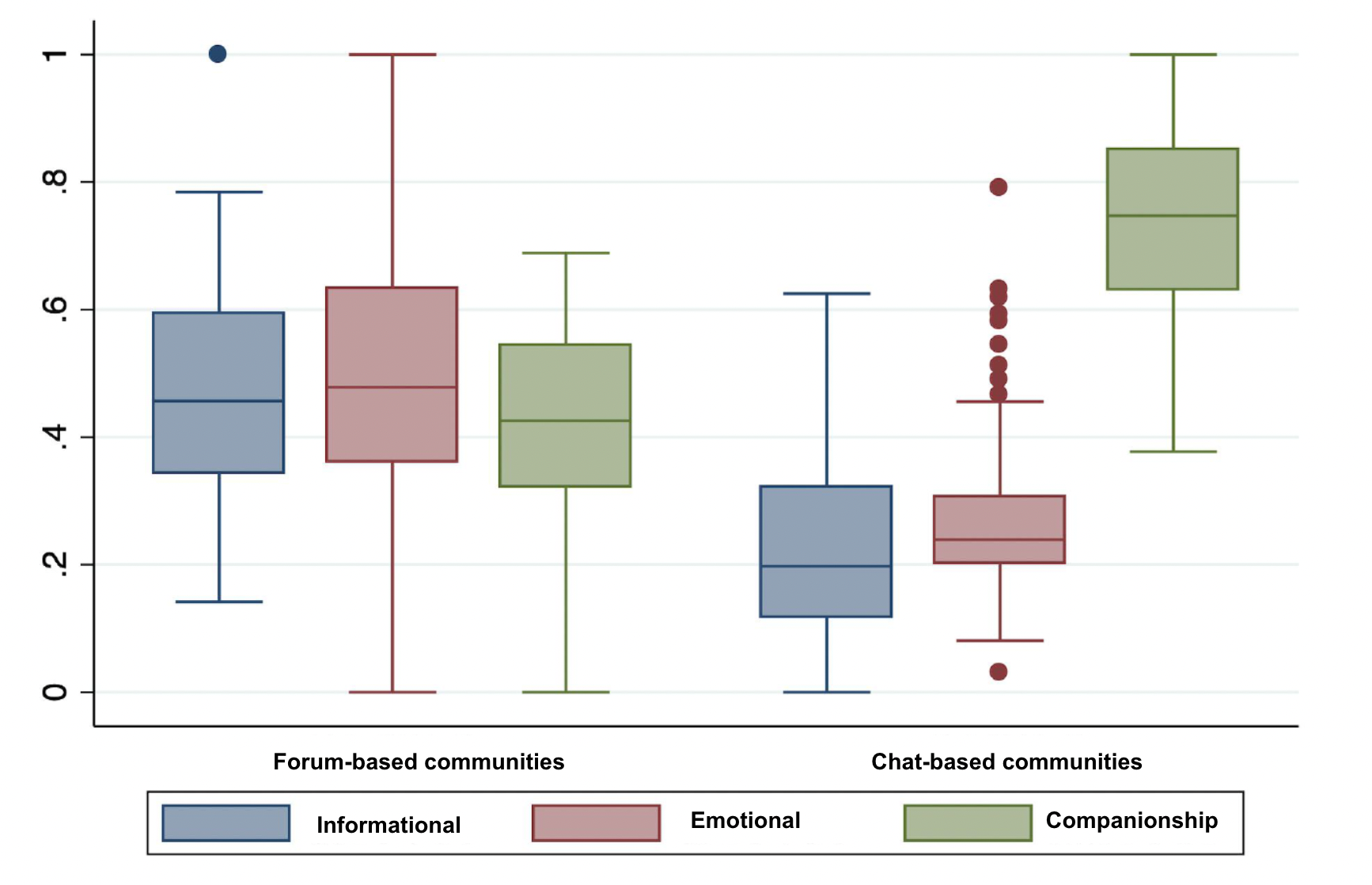}
  \caption{Box-plots of proportions of support types between forum-based and chat-based communities}
  \Description{}
  \label{fig2}
\end{figure}

The analysis revealed significant differences between forum-based and chat-based communities across all three categories of social support: informational support (z = 7.841, p < .001), emotional support (z = 6.737, p < .001), and companionship (t = –13.315, p < .001). As illustrated in Figure \ref{fig2}, forum-based communities exhibited higher proportions of informational and emotional support, whereas chat-based communities fostered greater levels of companionship. These findings suggest that the CMC format of a community is closely associated with the types of support exchanged among its members. Our results are consistent with prior work showing that discussion boards are more often used for informational exchanges, while blog and notes are conducive to the maintenance of interpersonal ties \cite{chuang_2014,Chuang_2012}. Moreover, research comparing online discussion forums and mobile instant messaging in collaborative learning contexts has found that forums tend to elicit more communication aimed at knowledge construction, whereas instant-messaging tools give rise to more socially oriented interactions \cite{sun_2017}. \cite{mansour_2024} further shows that chat rooms are frequently experienced as safe spaces for sharing "off-topic thoughts" and taking a "mental break" from task-focused learning discussions. Given these affordance differences across platforms, it is reasonable to interpret our findings as reflecting that asynchronous, threaded communication better supports reflective, problem-focused exchange — where users can take time to search for, compose, and archive information and resources for others — thus facilitating more help-focused support. By contrast, the real-time, conversational nature of chat communities appears to enable more causal, "off-topic" interactions for companionship within mental health support groups.   

\subsection{Differences in Communication Network Structure across Forum-based and Chat-based Communities}
\begin{figure}[!ht]
  \centering
  \includegraphics[width=\linewidth]{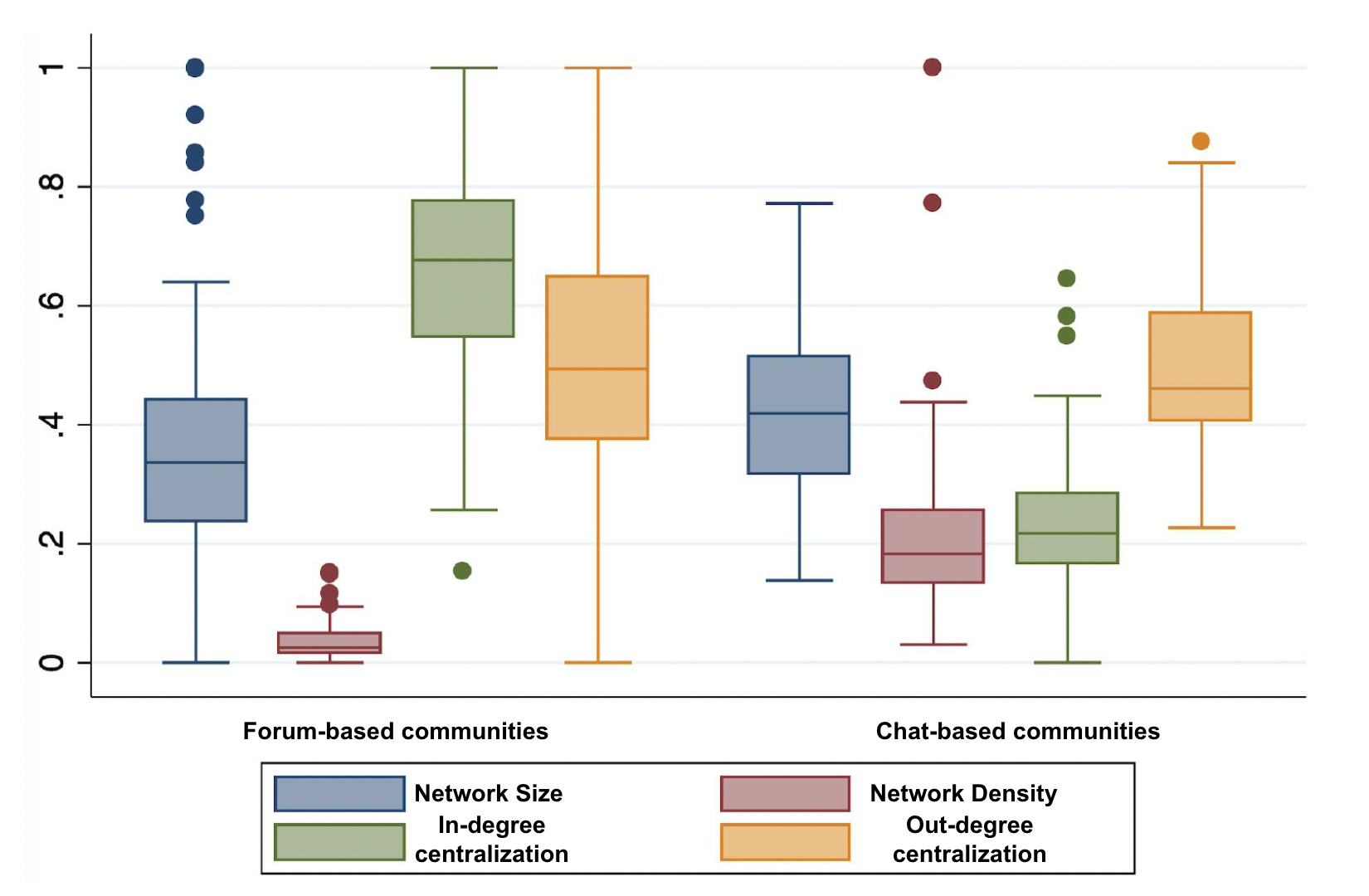}
  \caption{Box-plots of four structural characteristics of communication networks between forum-based and chat-based communities}
  \Description{}
  \label{fig3}
\end{figure}

\begin{figure}[!ht]
  \centering
  \includegraphics[width=\linewidth]{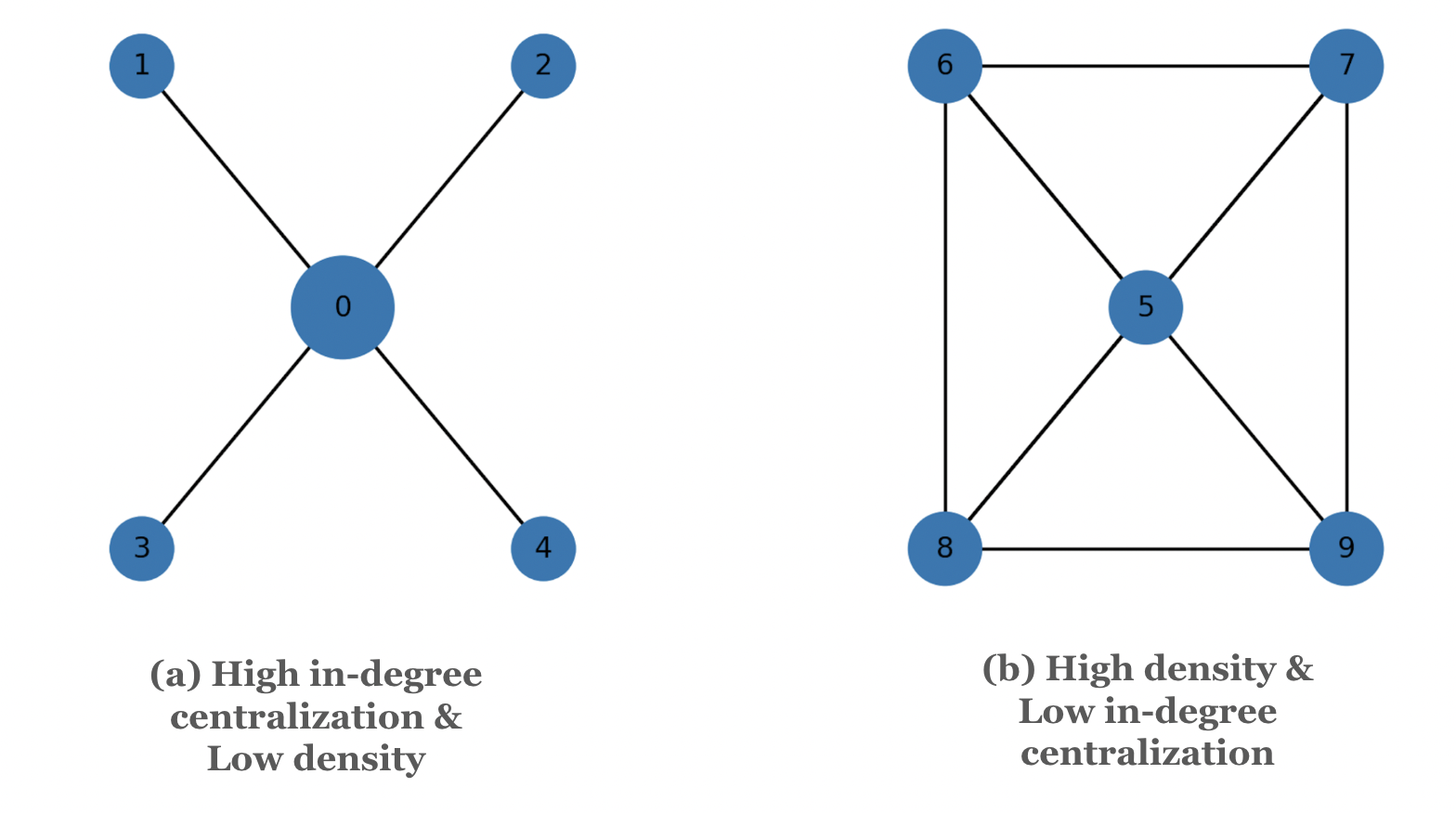}
  \caption{Illustrative communication network structures in forum-based vs. chat-based communities}
  \label{fig5}
  \vspace{2mm}
  \begin{minipage}{0.9\linewidth}
    \footnotesize \textit{Note.} This figure is a schematic illustration of different network structures and is not generated from the empirical data.
  \end{minipage}
  \Description{}
\end{figure}

To examine structural differences in the communication networks of forum-based versus chat-based online communities, we conducted independent samples t-tests and Wilcoxon rank-sum tests based on the results of normality checks. The analysis revealed significant differences between the two formats of community in network density (z = -10.275, p < .001) and in-degree centralization (z = 10.166, p < .001). However, no significant differences were found in network size (z = -3.27, p = .11) or out-degree centralization (t = 0.093, p = .926). Figure \ref{fig3} presents box-plots comparing the four structural characteristics between forum-based and chat-based communities. As shown, forum-based communities exhibited higher in-degree centralization, indicating that replies were more concentrated on a small number of participants. In contrast, chat-based communities showed higher network density, reflecting a more interconnected structure in which participants engaged more evenly with one another. As shown in the illustrative Figure \ref{fig5}, forum-based communities resemble a more "hub-and-spoke" structure in which a few highly visible members receiving a disproportionate share of replies. Notably, these members are often thread initiators, who can occupy a central and more visible position in the community by virtual of starting new discussions and having their posts listed as thread headers. This organizational structure naturally channels attention and responses toward a small subset of users. Conversely, chat-based communities exhibit a more web-like structure, where replies are more evenly distributed across members rather than flowing primarily to a few focal users. This pattern could be attributed to the stream-like organization of messages in chat interfaces, which creates conditions for more diffuse exchanges.  

\subsection{The Role of Communication Network Structure}
To examine whether the structural characteristics of communication networks account for the relationship between community format and the proportion of support types, we constructed a series of multilevel structural equation models (SEMs). In each model, format of community (coded as 0=forum-based communities; 1=chat-based communities) was treated as a predictor, and average message length was included as a covariate. The four network structural characteristics (i.e. network size, density, out-degree centralization, and in-degree centralization) were modeled to capture the indirect associations between format of community and proportion of support type in three separate SEMs. Given the modest sample size (N = 160), we assessed model fit using the Standardized Root Mean Square Residual (SRMR), a fit index particularly suited for small-sample SEMs \cite{hu_1999}. All three models yielded SRMR values below the recommended threshold of 0.08, indicating acceptable fit.

Results from the SEM analyses revealed statistically significant indirect associations via network structure for two types of support: informational support and companionship. For informational support, the indirect association was statistically significant ($\beta$ = –.05, p = .005), accounting for 91\% of the total association between community format and proportion of informational support. After accounting for the network structural variables, the direct path from community format to informational support was no longer significant ($\beta$ = –.005, p = .912), a pattern consistent with full statistical mediation. Among the four structural features, in-degree centralization emerged as the sole significant predictor ($\beta$ = –.042, p = .005), suggesting that the extent to which replies are concentrated among a few recipients was closely associated with levels of informational support in the community. In particular, forum-based communities, characterized by higher in-degree centralization, appear to create conditions that are especially conducive to information-oriented exchanges.

For companionship, the total indirect association was also significant ($\beta$ = .056, p = .015), explaining approximately 37\% of the total association between community format and proportion of companionship. The inclusion of network structural variables reduced the direct path from community format to companionship to a marginal level of significance ($\beta$ = .095, p = .071), indicating partial statistical mediation. Again, in-degree centralization was the only significant indirect pathway ($\beta$ = .039, p = .05), suggesting that lower concentration of replies — a hallmark of chat-based communities — fosters a more relational, companionship-oriented mode of interaction.

In contrast, the mediation model for emotional support was not supported. Although network density showed a small but statistically significant indirect association ($\beta$ = –.011, p = .003), the total indirect association was not significant, and the direct path from community format to emotional support remained robust ($\beta$ = –.087, p < .001), accounting for 91\% of the total association. These results suggest that the exchange of emotional support may be tied to platform-specific affordances or social norms that are not captured by network structure alone, and thus warrant investigation in future research.  

\section{Discussion}
\subsection{A Technical-Structural-Functional Model of Social Support} 
Grounded in the structural–functional model of support, the present study extends the theory and introduces the \textit{technical-structural-functional model of social support }within the context of computer-mediated communication. This model conceptualizes online communities as socio-technical systems, in which the medium that enables and sustains the community also critically shapes the interactions unfold within it. To empirically validate this framework, we focus on two distinct and prominent formats of online peer support communities—forum-based and chat-based — and examine both the structural characteristics of their communication networks and types of social support members exchanged. By assessing the mediating role of communication network structures, this approach provides a potential explanatory mechanism to explain the systematic differences in the nature of support across different platforms. 

\textbf{Technical Structures Shape Support Types. }The first research question asks whether, and how, the types of social support differ between forum-based and chat-based communities. Results show that community format significantly shapes the distribution of informational, emotional, and companionship support. Although both forum-based and chat-based communities exhibited the overall patterns with companionship being most prevalent, followed by informational support, and emotional support, the relevant prevalence of the three support types differ substantially. Compared with chat-based communities, forum-based communities generated approximately 16.3\% more informational support and 7.5\% more emotional support, whereas chat-based communities exhibited 23.8\% more companionship support. 

Our findings indicate that forum-based communities more often fulfill the help seeking and providing functions of social ties characterized by higher levels of informational and emotional support, whereas chat-based communities are more likely to facilitate hedonic forms of support in the form of companionship. While companionship is largely oriented toward intrinsic enjoyment instead of problem coping, prior research has demonstrated that it nonetheless produces tangible benefits for individual well-being \cite{rook_1987}. The types of support exchanged in peer support environments serve distinct member needs. Informational support involves the provision of guidance, advice, and resources directly relevant to managing mental health conditions. Emotional support consists of expressions of empathy, comfort, and validation, grounded in members’ lived experiences of mental health challenges. Companionship, by contrast, encompasses socially engaging interactions that may not be explicitly tied to mental health topics, yet still offer a sense of social presence, belonging and enjoyment. 

From this perspective, forum-based communities resemble Q\&A hubs, where individuals facing stressors or specific problems seek targeted informational and emotional assistance to address immediate concerns. However, once these needs are satisfied, participants may disengage, which is consistent with prior evidence that forum-based platforms often experience low member retention rates \cite{li_2021}. In contrast, chat-based communities function more like "social lounges". While members may initially join due to shared mental health experiences, the real-time, conversational affordances of chat platforms promote more fluid, informal exchanges. Over time, these interactions could expand beyond strictly mental health-related topics, sustaining everyday social connection and long-term companionship. 

\textbf{Technical Structures Shape Communication Network Structures. } With respect to communication networks, we found that forum-based communities exhibited higher in-degree centralization, whereas chat-based communities displayed higher network density. This pattern aligns with prior research noting denser interaction networks in chatroom environments \cite{owen_2016,zhao_2016}. The greater in-degree centralization in forum-based communities indicates that a small number of core members received disproportionately more replies than the majority of participants. Given the structural differences in message organization between the two community formats, it is plausible that, in forum-based communities, thread initiators attract greater attention and become the primary targets of replies. To test this assumption, we conducted a follow-up analysis comparing the number of messages received by thread initiators and non-initiators across the two community formats. In chat-based communities, initiators received twice as many replies as non-initiators. In contrast, in forum-based communities, initiators received 5 to 10 times as many replies. 

A comparison of the network sizes across the 20 forum-based and 20 chat-based communities revealed a counterintuitive pattern: chat-based communities exhibited a slightly larger average network size (M = 4.29) than forum-based communities (M = 4.06). This finding is particularly noteworthy given that forum-based communities generally have a much larger total membership base and are more open and accessible, making it easier for new members to join compared to the more restricted membership size that characterizes many chat-based groups.  

One plausible explanation lies in the immediacy and notification mechanisms embedded in chat platforms. The synchronous nature of chat-based interactions, combined with real-time alerts or push notifications, enables members to respond quickly, remain aware of ongoing discussions, and be drawn into conversations as they unfold. This immediacy not only encourages rapid exchanges but also lowers the activation threshold for participation, leading to a higher proportion of members actively engaging in dialogue at any given time. The asynchronous structure of forums, however, while allowing for more well-crafted and persistent messages, may inadvertently dampen ongoing engagement. As a result, even though forums have a larger potential pool of participants, the proportion of members actively contributing at any one time may be lower, yielding smaller active network sizes compared to their chat-based counterparts. It is also important to note that, given our one-month observation window, we interpret this finding with caution. Participation in forum-based communities often unfolds over longer periods, as one threads span months or even years and gradually expand the pool of active participants.

\textbf{Communication Network Structures and Support Dynamics across Platforms. }An examination of the mediating effects of the four network features revealed that the structure of communication networks plays a significant role in explaining the relationship between community format and the distribution of social support. Among these features, in-degree centralization emerged as the most influential indicator, though it shaped support dynamics in different ways across community formats. In forum-based communities, higher in-degree centralization means that a small subset of members—often thread initiators—receive a disproportionate share of replies, concentrating informational exchanges around these highly visible “hubs.” This pattern aligns with the observation that fact-oriented discussions tend to generate more centralized and hierarchical interaction structures than opinion-oriented discussions, which are typically more evenly distributed \cite{himelboim_2008}.

The relatively decentralized message-receiving networks in chat-based communities help account for their higher proportion of companionship support. Here, interaction patterns are more egalitarian, giving members roughly equal chances of receiving replies and encouraging socially rewarding, reciprocal exchanges. Prior research shows that receiving a reply signals attention from others and increases participants’ willingness to remain active in the community \cite{arguello_2006}. In Computer-Mediated Communication, response immediacy serves as a salient nonverbal cue of co-presence, and perceived social presence strongly shapes how people evaluate and relate to interaction partners \cite{walther_1995}. The high synchronicity and interactivity of chat systems, coupled with more balanced reply patterns, thus likely heighten social presence and encourage sustained engagement. Studies of immersive platforms, such as cancer support groups in Second Life, have shown that members often describe fellow participants as “friends” and feel as though they are “together in the same place” \cite{green_hamann_2011}. These dynamics help explain why, despite their smaller total membership, chat-based communities can maintain active network sizes comparable to those of forum-based communities, and why companionship-oriented exchanges become a defining mode of support in these environments.

\subsection{Design and Practical Implications}
This study offers important practical contributions for the technical development and design of online mental health support communities. Building on our findings, we argue that the organization of messages and the temporal structure of communication are key determinants of how members interact and which forms of social support predominate.

From a \textit{technical design perspective}, forum-based communities — characterized by asynchronous, persistent, and hierarchically organized discussions — tend to generate highly centralized message-receiving networks. Such a structure is advantageous for targeted informational exchanges, particularly for a small number of central members who act as thread initiators. However, this centralization also limits opportunities for less prominent members to receive replies under the initiators' threads, potentially discouraging their continued participation. To address this, designers of forum-based systems could implement interface features that promote more distributed and reciprocal engagement. 

\begin{designbox}
\textbf{Design Recommendation 1: From Central Voices to Shared Visibility in Forums}
\begin{itemize}
    \item Introduce an \emph{“unanswered posts” sidebar} that automatically highlights threads with few or no replies and prompts members to respond;
    \item Implement a \emph{“random spotlight”} module that periodically surfaces under-replied or newcomer messages to diversify whose voices are seen, regardless of whether they appear as thread-initiating posts;
    \item Adjust default \emph{sorting and recommendation algorithms} so that some portion of the feed is reserved for low-visibility threads or comments, rather than ranking posts solely by popularity or recency;
\item Provide \emph{configurable notification settings} that allow users to opt in to alerts about “posts that still need a reply” in topics they care about, distributing attention in a more user-centric manner.
\end{itemize}
\end{designbox}

These features can redistribute attention, reduce reply concentration, and create more opportunities for peripheral members to receive responses, while still preserving the advantages of structured, persistent discussions.

By contrast, chat-based communities — marked by high synchronicity and relatively flat message organization — tend to produce more decentralized communication networks. Such egalitarian patterns of interaction give members roughly equal chances of receiving replies, fostering socially rewarding, companionship-oriented exchanges. The high degree of social presence created by synchronous messaging, combined with frequent reciprocal interactions, can strengthen interpersonal bonds and sustain participation. Yet the same immediacy that drives engagement can make it challenging to locate or revisit useful information and resources once it scrolls out of view.  To address this tension, designers of chat-based platforms could:

\begin{designbox}
\textbf{Design Recommendation 2: Making Information Findable in Chat Groups}
\begin{itemize}
    \item Integrate a persistent \emph{“knowledge board”} that aggregates and pins helpful informational posts (e.g., treatment options, crisis resources) so they do not get lost in the chat stream;
    \item Allow messages in the chat to be \emph{tagged and saved into topic-specific collections}, making it easier to browse relevant advice without leaving the chat space;
    \item Provide \emph{searchable archives} with simple filters so members can quickly retrieve prior informational exchanges when similar questions arise.
\end{itemize}
\end{designbox}

These design elements preserve the socially rich, companionship-oriented nature of synchronous chats while making informational support more visible, organized, and reusable over time.

From a \textit{user perspective}, understanding these structural and functional differences can help individuals seeking social support make more informed choices about where to participate. Our findings suggest that: 

\begin{designbox}
\textbf{User Recommendation: Choosing Communities Based on Support Needs}
\begin{itemize}
    \item If you prioritize social support for coping specific problems, difficulties, or distress, forum-based communities may be more effective, as their persistent threads and more centralized reply patterns tend to concentrate targeted resources and make on-topic content easier to locate and revisit; 
    \item If you prioritize companionship and everyday connection (e.g., feeling less alone, casual conversation, ongoing check-ins), chat-based communities may be more satisfying, as synchronous, reciprocal exchanges and more egalitarian reply patterns provide frequent, low-barrier opportunities for "off-topic" interaction.
\end{itemize}
\end{designbox}

More broadly, these insights point to a \textit{structure-function-driven approach to }online peer support community design, in which message organization and synchronicity can be treated as adjustable levers to align platform affordances with desired support outcomes. Our findings also underscore the trade-offs between informational value and companionship value in designing peer support systems. Rather than assuming that a single format can fully meet all support needs, it may be more realistic to consider how different features can be combined to emphasize one function while still accommodating the other. In practice, relative small design changes can be explored to make user experiences smoother. For example, forum platforms such as Reddit have experimented with community chat features, which allow users to carve out spaces for more real-time interactions \footnote{\url{https://redditforcommunity.com/features/chat-channels}}. Group chat platforms like Discord, on the other hand, have introduced features such as "thread" that group messages into topic-based subchannels, making it easier to preserve and revisit useful exchanges \footnote{\url{https://discord.com/developers/docs/topics/threads}}. These kind of hybrid arrangements suggest that adjusgments to message organization and synchronicity can be feasible, and help navigate, rather than eliminate the tension between different types of platforms.

\subsection{Limitations and Future Work}
This study has several limitations that should be considered when interpreting its findings. First, although we identified associations between community format, network structure, and social support type, the data and analytical approach do not permit causal inferences. Other unmeasured factors — such as community norms, member roles, or additional network features — may also influence these relationships. Future research should draw on relevant theoretical frameworks to examine other socio-technical mechanisms underlying systematic variations in social support across online communities, and employ experimental or quasi-experimental designs to test causal links.

User self-selection is likely an important explanatory factor, however, we cannot fully disentangle platform affordances from self-selection effects. Our analyses compare community-level structures and support patterns across forum-based and chat-based platforms, but we do not have detailed data on users' demographics, mental health conditions, or initial motivations for joining each community. It is therefore plausible that individuals who seek informational support deliberately gravitate toward online forums, while those who seek companionship join chat groups. At the same time, our findings do not exclude the possibility that people can and sometimes do choose spaces that partially match their needs and are subsequently shaped by the communication patterns they encounter there. In reality, platform affordances and user motivation and characteristics are likely to interact. Future research using mixed-method approaches linking community-level data to user-level data will be important for unpacking how technical and social factors jointly contribute to the observed differences in the nature of support exchanged across different online platforms.

Second, our analysis focused on online support communities centered on various mental health conditions. While this focus is practically significant, given the growing prevalence of mental health challenges and the importance of understanding the support needs and interactions of those affected, there is also a need to investigate communities devoted to other major health issues (e.g., cancer, alcohol misuse, obesity) as well as political or social topics. Because a community’s focal theme can shape both its interaction structure and content \cite{himelboim_2008}, the generalizability of our findings beyond mental health contexts remains uncertain. Moreover, since we only collected data from four Chinese platforms that host online communities, the findings should be interpreted with caution when generalizing to other platforms or cultural contexts. Future research could test the proposed model across diverse cultural settings and across forum- or chat-based platforms with varying features to evaluate the robustness and boundary conditions of the framework.

Third, we adopted a "three-category" classification of social support—informational, emotional, and companionship—which, although commonly used, excludes other important forms such as esteem or network support, as well as non-supportive interactions. This supervised machine learning model allowed us to directly interpret which linguistic markers and topics were associated with different support types. We made this choice partly to prioritize interpretability and theoretical insight into how specific lexical, topical, and affective features map onto informational, emotional, and companionship support, which is more difficult to obtain from large, highly parameterized models. In addition, our data contain sensitive mental health disclosures, which makes it challenging to transmit raw text to proprietary large language models. Future work could examine data analytical approaches that leverage state-of-the-art transformer models or open-source large language models to better balance interpretability, performance, and data protection. Moreover, our study was constrained by a single-coding approach, in which each textual entry was annotated with only one primary support type for the sake of model performance.  This inevitably simplifies the complexity of real-world interactions, as a single message, especially longer ones, may convey multiple forms of support.

In addition, our study focuses on broader categories of informational, emotional, and companionship support across the two types of platforms; a more fine-grained analysis of support subtypes falls beyond the scope of the present work. It is undoubtedly possible that, even within the same overarching category (e.g., informational support), there are meaningful differences in subtypes and linguistic expressions across platforms. Such variation may be attributable to affordances — for instance, chat platforms may encourage shorter, more conversational and fragmented messages, whereas forums may facilitate longer and more elaborated messages \cite{rettie_2009,oztok_2013}. Future work could leverage large language models, which are better equipped to disentangle nuanced meanings within text, to implement fine-grained, multi-label or segment-level coding. Such approaches may more fully capture the layered nature of support exchanges and yield a richer understanding of how different support types co-occur within the same message. 

Finally, our dataset comprised one month of interactions from 40 online mental health support communities, with the time frame selected due to data access limitations. This design does not capture longer-term or historical changes in community dynamics. Prior research indicates that support content may evolve as members’ participation accumulates over time \cite{rains_2009}. Longitudinal analyses could therefore yield deeper insights into the temporal dynamics of social support exchange.

\bibliographystyle{ACM-Reference-Format}
\bibliography{sample-base}

\appendix
\section{Appendix}
\subsection{Coding Schema and Examples}
\begin{table}[H]
    \centering
    \begin{tabular}{c>{\centering\arraybackslash}p{100pt}>{\centering\arraybackslash}p{100pt}}\toprule
         Support Type&  Working Definition& Example\\\midrule
         Informational Support&  This refers to content related to information, advice, references, or guidance that facilitates problem solving. This may include discussions of medication, symptoms, side effects, treatment options, and similar topics.& "Quetiapine is a relatively effective medication, with higher-quality packaging in imported versions and comparatively mild side effects." \\
         Emotional Support&  This refers to content that related to the expressions of caring, concern, comfort, empathy, sympathy, and validation& "Stay strong and open your heart. We will all get better in time"\\
 Companionship Support& This refers to off-topic causal conversation unrelated to mental health, initiated for the purpose of pleasurable social interaction or relational connection.&"I prefer the version by Hiroshi Akita.  It's performed by the original songwriter of the piece."\\ \bottomrule
    \end{tabular}
    \caption{Coding schema and examples of user-generated content}
    \label{tab:placeholder}
\end{table}

\subsection{Confusion Matrix}
\begin{table}[H]
    \centering
    \begin{tabular}{ccccc}\toprule
         &  &  \multicolumn{3}{c}{Coder 2}\\\midrule
         &  &  Informational&  Emotional& Companionship\\
         &  Informational&  335&  5& 5\\
         Coder 1&  Emotional&  3&  325& 10\\
         &  Companionship&  8&  7& 302\\ \bottomrule
    \end{tabular}
    \caption{Confusion matrix for manual annotation of support type}
    \label{appendix_table_2}
\end{table}

\end{document}